\documentclass[a4paper,11pt]{article}
\usepackage{jinstpub} 
 
\usepackage{lineno}
\usepackage{array}
\usepackage{hyperref} 
\proceeding{The 7$^{\text{th}}$International Conference on Micro Pattern Gaseous Detectors\\
December 11-16, 2022.\\
Rehovot, Israel.}

\title{\boldmath A large area 100 channel Picosec Micromegas detector with sub 20 ps time resolution}

\author[a,1]{\footnotesize A. Utrobicic,\note{Corresponding author.}}
\author[b]{Y. Angelis}
\author[c]{J. Bortfeldt}
\author[d]{F. Brunbauer}
\author[b]{E. Chatzianagnostou}
\author[e]{K. Dehmelt}
\author[f]{G. Fanourakis}
\author[d,g]{K. J. Floethner}
\author[h]{M. Gallinaro}
\author[i]{F. Garcia}
\author[e]{P. Garg}
\author[j]{I. Giomataris}
\author[k]{K. Gnanvo}
\author[l]{T. Gustavsson}
\author[j,2]{F.J. Iguaz,\note{2Now at SOLEIL Synchrotron, L’Orme des Merisiers, Départementale 128, 91190 Saint-Aubin, France.}}
\author[a,m,n]{D. Janssens}
\author[j]{A. Kallitsopoulou}
\author[o]{M. Kovacic}
\author[j]{P. Legou}
\author[d,p]{M. Lisowska}
\author[q]{J. Liu}
\author[g,r]{M. Lupberger}
\author[k]{S. Malace}
\author[d,b,3]{I. Maniatis, \note{Now at Department of Particle Physics and Astronomy, Weizmann Institute of Science, Hrzl st. 234, Rehovot,
7610001, Israel.}}   
\author[q]{Y. Meng}
\author[d,r]{H. Muller}
\author[d]{E. Oliveri}
\author[d,s]{G. Orlandini}
\author[j]{T. Papaevangelou}
\author[t]{M. Pomorski}
\author[d]{L. Ropelewski}
\author[b,u]{D. Sampsonidis}
\author[d,r]{L. Scharenberg}
\author[d]{T. Schneider}
\author[j,4]{L. Sohl, \note{Now at TÜV NORD EnSys GmbH Co. KG.}}
\author[d]{M. van Stenis}
\author[v]{Y. Tsipolitis}
\author[b,u]{S.E. Tzamarias}
\author[d, w]{R. Veenhof}
\author[q]{X. Wang}
\author[d,x]{S. White}
\author[q]{Z. Zhang}
\author[q]{Y. Zhou}

\affiliation[a]{Ruđer  Bošković Institute, Bijeni\v{c}ka cesta 54, 10000, Zagreb, Croatia}

\affiliation[b]{Department of Physics, Aristotle University of Thessaloniki, University Campus, GR-54124, Thessaloniki, Greece}

\affiliation[c]{Department for Medical Physics, Ludwig Maximilian University of Munich,  Am Coulombwall 1, 85748 Garching, Germany}

\affiliation[d]{European Organization for Nuclear Research (CERN), CH-1211, Geneve 23, Switzerland}

\affiliation[e]{Stony Brook University, Dept. of Physics and Astronomy, Stony Brook, NY 11794-3800, USA}

\affiliation[f]{Institute of Nuclear and Particle Physics, NCSR Demokritos, GR-15341 Agia Paraskevi, Attiki, Greece}

\affiliation[g]{Helmholtz-Institut für Strahlen- und Kernphysik, University of Bonn, Nußallee 14–16, 53115 Bonn, Germany}

\affiliation[h]{Laboratório de Instrumentacão e Física Experimental de Partículas, Lisbon, Portugal}

\affiliation[i]{Helsinki Institute of Physics, University of Helsinki, FI-00014 Helsinki, Finland}

\affiliation[j]{IRFU, CEA, Université Paris-Saclay, F-91191 Gif-sur-Yvette, France}

\affiliation[k]{Jefferson Lab, 12000 Jefferson Avenue, Newport News, VA 23606, USA}

\affiliation[l]{LIDYL, CEA, CNRS, Universit Paris-Saclay, F-91191 Gif-sur-Yvette, France}

\affiliation[m]{Inter-University Institute for High Energies (IIHE), Belgium}

\affiliation[n]{Vrije Universiteit Brussel, Pleinlaan 2, 1050 Brussels, Belgium}

\affiliation[o]{Faculty of Electrical Engineering and Computing, University of Zagreb, 10000 Zagreb, Croatia}

\affiliation[p]{Université Paris-Saclay, F-91191 Gif-sur-Yvette, France}

\affiliation[q]{State Key Laboratory of Particle Detection and Electronics, University of Science and Technology of China, Hefei 230026, China}

\affiliation[r]{Physikalisches Institut, University of Bonn, Nußallee 12, 53115 Bonn, Germany}

\affiliation[s]{Friedrich-Alexander-Universität Erlangen-Nürnberg, Schloßplatz 4, 91054 Erlangen, Germany}

\affiliation[t]{CEA-LIST, Diamond Sensors Laboratory, CEA Saclay, F-91191 Gif-sur-Yvette, France}

\affiliation[u]{Center for Interdisciplinary Research and Innovation (CIRI-AUTH), Thessaloniki 57001, Greece}

\affiliation[v]{National Technical University of Athens, Athens, Greece}

\affiliation[w]{Bursa Uludaǧ University, Görükle Kampusu, 16059 Niufer/Bursa, Turkey}

\affiliation[x]{University of Virginia, USA}

\emailAdd{antonija.utrobicic@irb.hr }

\abstract{

The PICOSEC Micromegas precise timing detector is based on a Cherenkov radiator coupled to a semi-transparent photocathode and a Micromegas amplification structure. The first proof of concept single-channel small area prototype was able to achieve time resolution below 25 ps. One of the crucial aspects in the development of the precise timing gaseous detectors applicable in high-energy physics experiments is a modular design that enables large area coverage. The first 19-channel multi-pad prototype with an active area of approximately 10 cm$^2$ suffered from degraded timing resolution due to the non-uniformity of the preamplification gap. A new 100 cm$^2$ detector module with 100 channels based on a rigid hybrid ceramic/FR4 Micromegas board for improved drift gap uniformity was developed. Initial measurements with 80 GeV/c muons showed improvements in timing response over measured pads and a time resolution below 25 ps. More recent measurements with a new thinner drift gap detector module and newly developed RF pulse amplifiers show that the resolution can be enhanced to a level of 17~ps. This work will present the development of the detector from structural simulations, design, and beam test commissioning with a focus on the timing performance of a thinner drift gap detector module in combination with new electronics using an automated timing scan method. 
}


\begin{document}
\maketitle
\flushbottom

\section{Introduction}
\label{sec:intro}
One of the many challenges for future physics experiments is the development of precise timing detectors ($\mathcal{O}$ 100 ps) with a large area coverage and capabilities of withstanding high radiation fluence \cite{colaleo20212021}.
 PICOSEC Micromegas (MM) is a gaseous detector for precise timing applications with a time resolution of tens of ps \cite{bortfeldt2018picosec}. It consists of a MM amplification structure that is coupled to a Cherenkov radiator with a CsI semi-transparent photo-cathode. The detector is operating as a two-stage amplification structure with a working gas mixture of Ne:C$_2$H$_6$:CF$_4$ (80:10:10). Operating principles are in detail described in \citep{bortfeldt2018picosec, aune2021timing}. A single channel prototype achieved excellent time resolution below 25~ps for measurements with Minimum Ionizing Particles (MIPs) \citep{bortfeldt2018picosec} and developments continued towards a multi-channel detector that can cover larger areas \citep{aune2021timing}. An improved detector prototype with 19 hexagonal pads was built and tested. It was observed that this prototype suffered from the drift field non-uniformity. The effect was investigated and flatness corrections were made based on a particle hit position to obtain a uniform time response. After applying the necessary corrections, the detector yielded a time resolution of 25 ps in the best case \cite{aune2021timing}. It was found that the MM Printed Circuit Board (PCB) had initial deformations, most likely caused by the production process and the stretching of the mesh, which can be exacerbated by the tightening of the PCB to the detector housing. In order to produce an even larger multipad PICOSEC, new design procedures had to be introduced.
 
\section{Mechanical aspects and timing performance of the 100 channel prototype}
\label{sec:2}

\begin{figure}[!htb]
\begin{center}
\includegraphics[width=0.4\columnwidth]{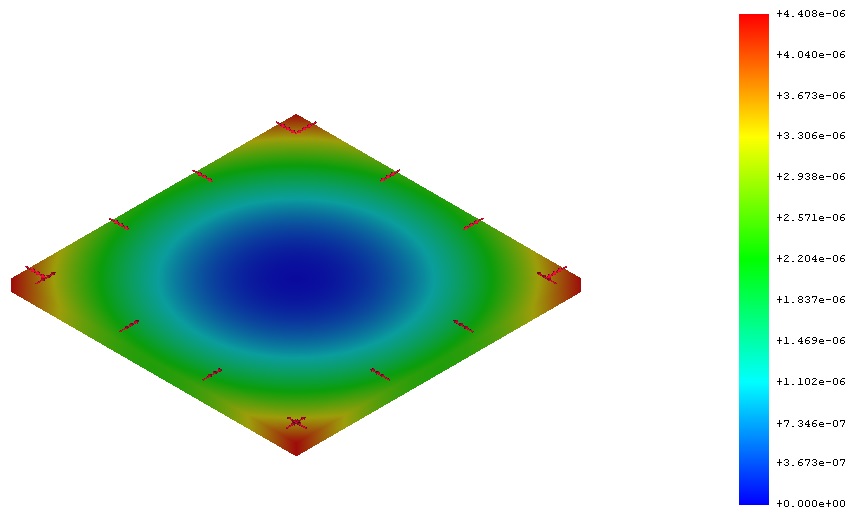}
\includegraphics[width=0.4\columnwidth]{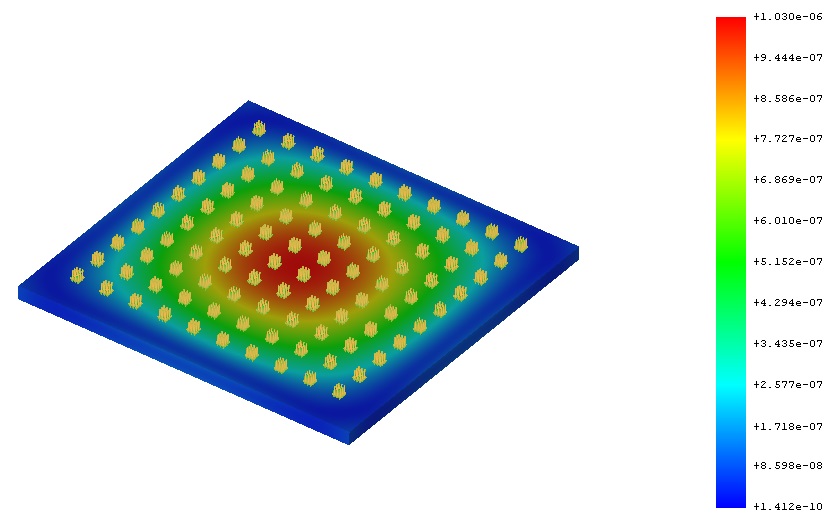}

\caption{Left: Simulation results of the mesh tension influence on 4~mm thick ceramic. A shear force of 15 N/cm was applied on the coverlay faces directed inward the active area. Right: Simulation result of the spring-loaded pin pressure. A Force of 23 g was applied to 100 pressing points. Ceramics Young's modulus was 320~GPa and the Poisson ratio was 0.23.}
\label{fig_mechanics}
\end{center}
\end{figure}

The design of the new 100-ch prototype was driven with improvements in the flatness and uniformity of the drift gap as the main objectives. Structural mechanics simulations of the mesh tension interface and coupling to the detector casing were performed to select the substrate that could ensure planarity better than 10~$\mu$m over 100 cm$^2$ active area, Figure \ref{fig_mechanics}. The Finite Element Analysis suggested the use of a more rigid and thicker PCB. The MM board was fabricated as a combination of 4~x~1~mm thick ceramic core with a 0.3~mm thick FR4 outer layers \cite{utrobicic_mechanics}. The ceramic core was used to achieve rigidity, while the use of the thin FR4 layers enabled the classic PCB manufacturing process, precise flattening, and polishing.
The readout structure was arranged in a 10 by 10 matrix of square pads. 
 
\begin{figure}[!htb]
\begin{center}
\includegraphics[width=0.4\columnwidth]{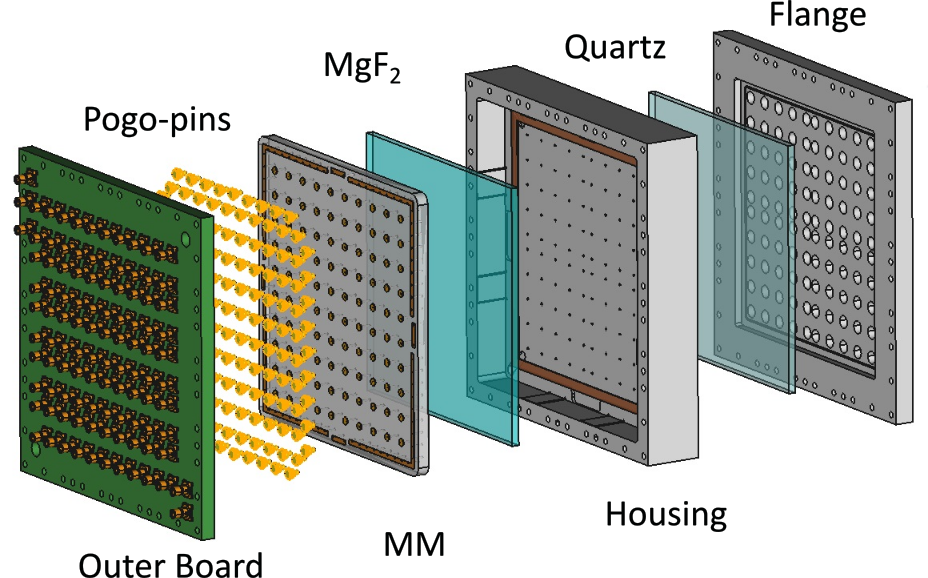}
\hspace{0.5cm}
\includegraphics[width=0.27\columnwidth]{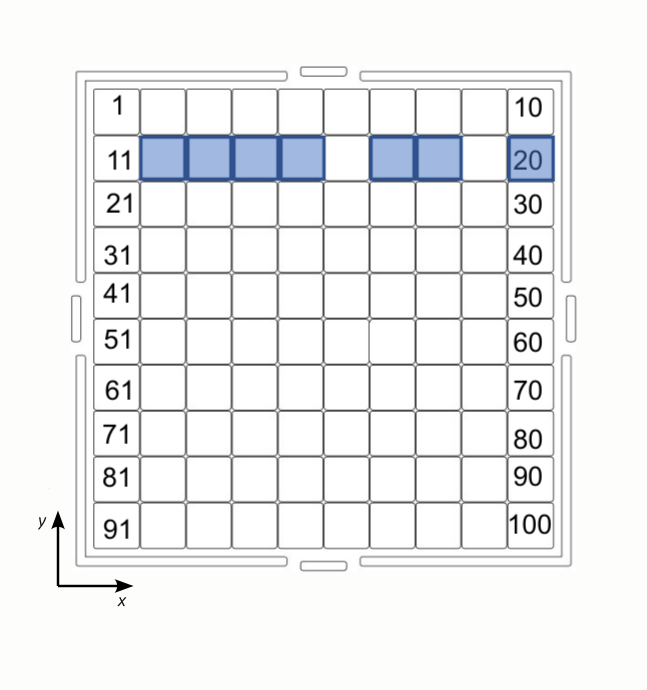}

\caption{Left: Exploded view of the detector layout. Right: MM PCB pad layout. Each pad is 9.82 mm x 9.82 mm with 200 $\mu$m spacing between the edges of adjacent pads. The blue squares represent the pads used for evaluating the timing performance over the detector area.}

\label{fig_layout}
\end{center}
\end{figure}

To prevent bending of the MM board or the flange with a Cherenkov radiator when pressed with screws, it was crucial to mechanically detach all the active detector components, including the MM board and the Cherenkov radiator, from the housing. The design choice was to use an Outer Board (OB) on the side of the chamber where the MM board is located, and a flange with a quartz window on the other side where the Cerenkov radiator is located, Figure \ref{fig_layout} (left). The signal from each pad was routed to the SMB connectors on the top of the OB via spring-loaded pins. In addition to the electrical connection, the compressive force of the spring-loaded pins ensures that the radiator is pressed firmly and evenly into the positioning groove in the chamber.

The time response of the 100-channel Detector Under Test (DUT) to 80~GeV muons was measured at the CERN SPS H4 secondary beamline. The setup consisted of a triple GEM detector telescope for a particle hit position information and MCP-PMTs\footnote{Hamamatsu MCP-PMT R3809U-50 \url{https://www.hamamatsu.com/jp/en/product/type/R3809U-50/index.html}} (later in the text MCP) used as a timing reference or trigger \citep{aune2021timing, bortfeldt2018picosec}. In a standard readout chain configuration, an induced signal from a detector is amplified using high bandwidth CIVIDEC C2\footnote{C2-HV Broadband amplifier (2 GHz, 40 dB), \url{https://cividec.at/electronics-C2-HV.html}} amplifier and digitized at a sampling rate of 10 GS/s by LeCroy WR8104 oscilloscope. The same oscilloscope is used for the acquisition of the reference time detector signal and serial bitstream with event ID obtained from the tracker. The recorded waveforms are analyzed using the methods described in previous publications \citep{bortfeldt2018picosec, aune2021timing}.

The first measurements were made with a prototype detector with a "standard" drift gap thickness of 220~$\mu$m and a CsI photocathode on a thin Cr layer. Two MCP detectors were aligned with each other in the x-y plane and the DUT was positioned behind them. To be able to take data from particles crossing an entire pad area, one MCP was used as a trigger detector. It was mounted parallel to the DUT and placed in front of it. In addition, another MCP, aligned with the trigger MCP in the x-y plane, was placed in front of it and used as a timing reference.
For the timing measurements, the DUT was fixed so that the measured pad (or detector area) was aligned with the trigger and timing MCP. Measurements of more than 10 pads showed that a time resolution below ~25~ps can be obtained for each pad in its central square area of 5 mm x 5 mm, thus reproducing timing properties of the single channel prototype. 

\begin{figure}[!htb]
\begin{center}
\includegraphics[width=0.3\columnwidth]{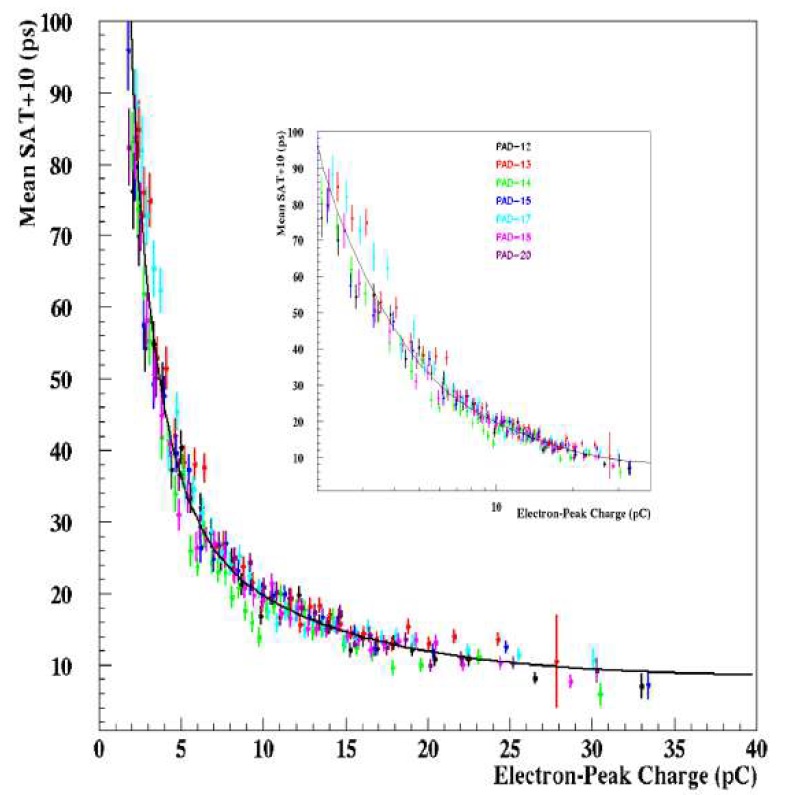}
\hspace{0.5 cm}
\includegraphics[width=0.3\columnwidth]{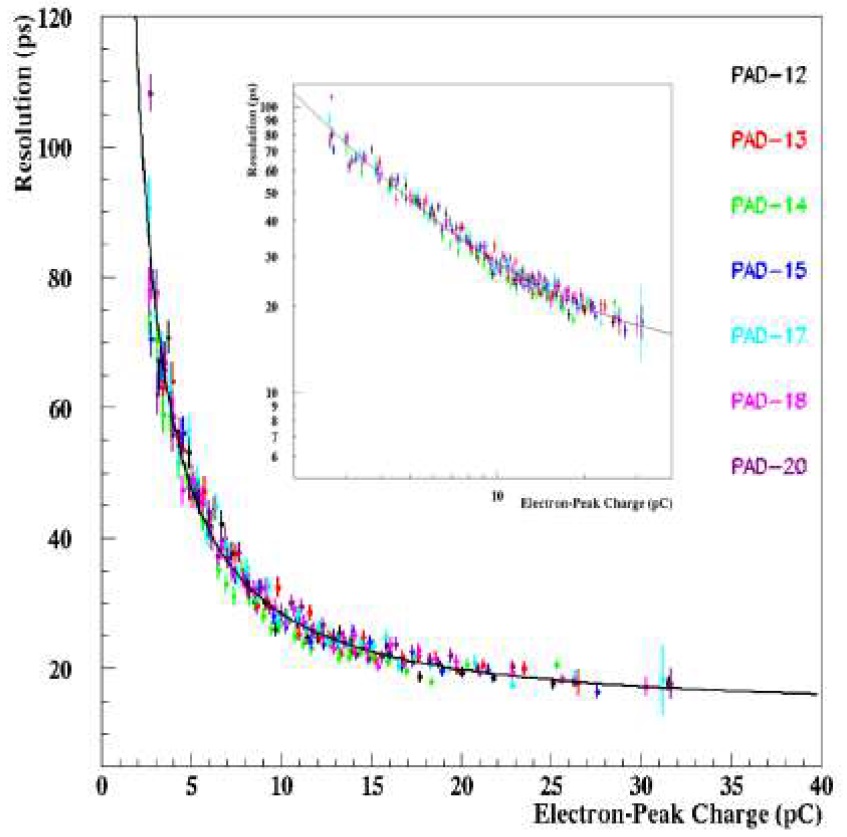}
\caption{Mean SAT (left) and time resolution (right) values for different electron peak charge bins \cite{maniatis2022research}.}
\label{fig_timing220}
\end{center}
\end{figure}

The timing of the detector can be further improved if the dependence of the signal arrival time (SAT) on the electron peak (e-peak) charge is taken into account as a time-walk correction. If the drift field is uniform over the entire surface of the pad, the dependence of the timing characteristics on the e-peak charge should be the same for all pads. One of the most important requirements for a large area PICOSEC MM detector is that all pads have a similar dependence of the mean SAT and time resolution on the e-peak charge. This would suggest that it is possible to use only a single global time walk correction function for all the pads, making it easier to calibrate the detector. To investigate the timing performance over a larger area, timing measurements were taken from seven different pads in the same row, Figure \ref{fig_layout} (right). The timing analysis shows that all pads closely follow the time-walk correction function, indicating a consistent global behaviour that is important for many-channel detectors, Figure \ref{fig_timing220} \citep{maniatis2022research}.

\section{Method for timing characterization of large area detectors}

Measuring the timing performance of large-area detectors presents a challenge for the selection of the reference detector, which must have better timing characteristics and spatial uniformity than the DUT. Although large-area MCPs are available, they are more expensive and the timing performance over the entire area must be known to ensure that the DUT response measurements are not biased. In the previous work \cite{sohlspatial, bortfeldt2020timing}, the timing of the reference detector MCP was investigated as a function of the particle hit position at different radii from the centre of the detector. It was observed that the MCP can be used as a timing reference over the effective photocathode diameter (11~mm), with a time resolution below 6~ps. In the outer regions, the time resolution deteriorates rapidly. To conduct a measurement with minimal bias caused by the non-uniformity of the reference detector over larger DUT areas (> 11~mm), a modification to the measurement method had to be made.

\begin{figure}[!htb]
\begin{center}
\includegraphics[width=0.3\columnwidth]{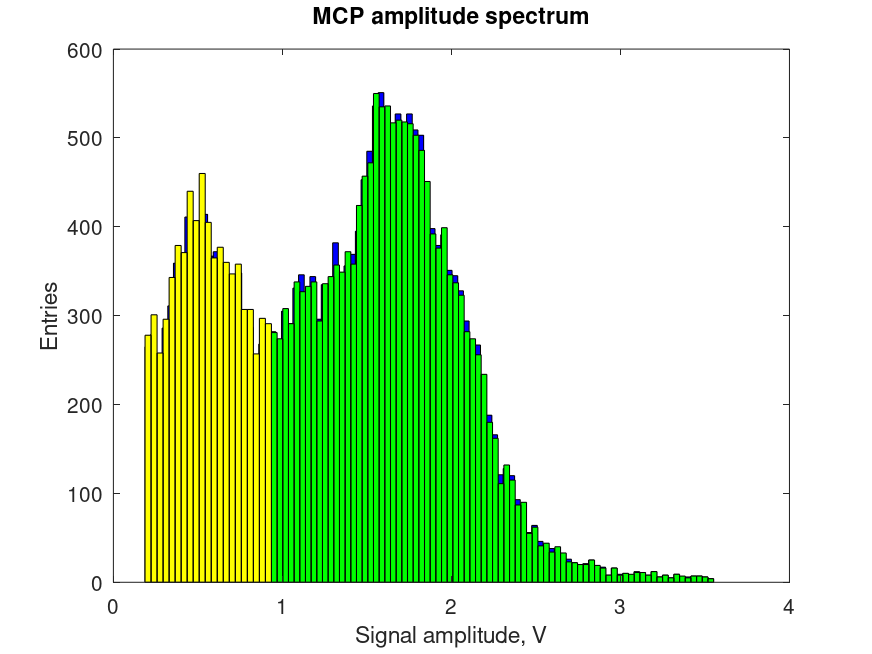}
\includegraphics[width=0.4\columnwidth]{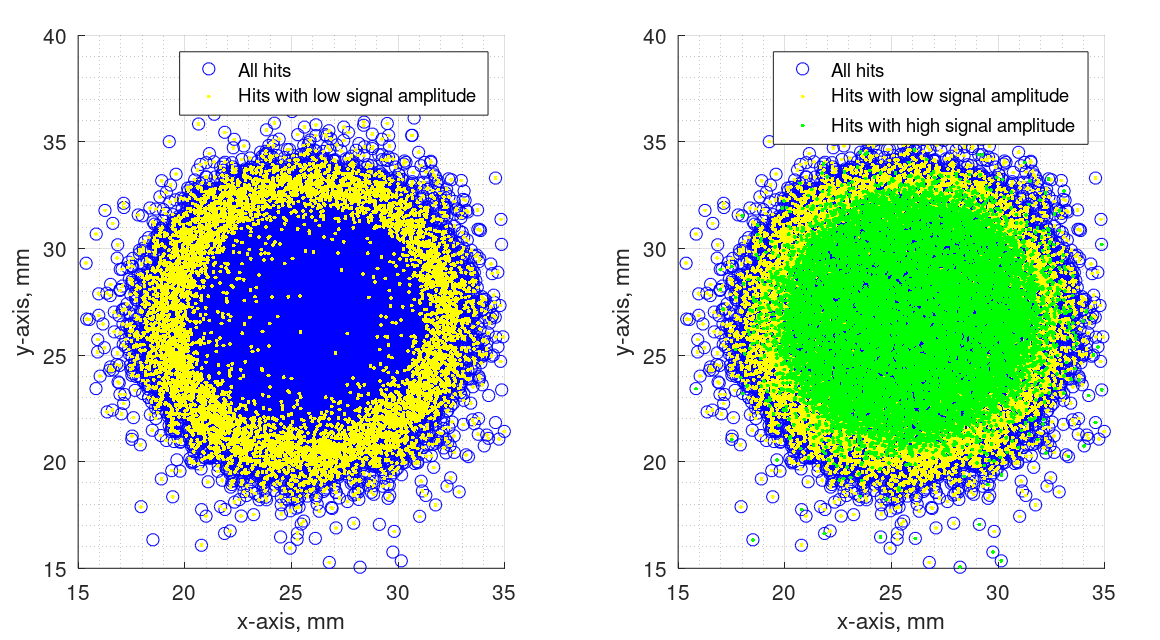}
\caption{Left: Spectrum of the MCP amplitudes (below 1~V in yellow and above 1~V in green). Middle: Hit map of the lower amplitudes (yellow). Right: Hit map of the higher amplitudes (green). }  
\label{fig_mcp_spectrum}
\end{center}
\end{figure}

The analysis of the measurements at the reference MCP taken with MIPs shows a double-peak amplitude spectrum, Figure \ref{fig_mcp_spectrum} (left). The correlation of the amplitude spectrum with the hit positions obtained using the tracker shows that the events in the lower signal amplitude peak belong to the outer region of MCP (Figure \ref{fig_mcp_spectrum}, middle), while those in the larger amplitude peak correspond to the hits in the center of the MCP,  Figure \ref{fig_mcp_spectrum} (left). The method is based on using the central ("good") part of the MCP to perform timing measurements. This was realized by using the same MCP as a trigger and as a time reference with a sufficiently high threshold to reject the events from the outer region. The MCP was mounted to the movable stage and an automated scanning method was employed to cover the large DUT area with the central part of MCP. The movement was synchronised with the number of events recorded to ensure an even amount of data over the entire area. The MCP was moved on a grid of 2.5~mm x 2.5~mm every 1000 recorded events. A dense grid of movement ensured an overlap of about 75\ \%, which is beneficial to average possible imperfections of the MCP. Although the method is time-consuming due to the geometric scan, it benefits from the reduced load on the tracking system and oscilloscope recording. This approach can be employed for measurements on a single pad, signal sharing, or entire detector scan studies.

\section{Beam test measurements with custom electronics and 180~$\mu$m drift gap thickness}
Earlier measurements and simulations indicated that the reduction of the drift gap improves the time resolution of the PICOSEC MM \cite{bortfeldt2021modeling, sohl2020single}. After the initial tests described in section \ref{sec:2} showed the preservation of the timing performance of the single-channel detector and confirmed the uniformity of the detector, the drift gap was reduced to 180~$\mu$m. In addition, 50 channels of the detector were fitted with custom-made 10-channel, 38~dB, 650~MHz amplifier cards with built-in discharge protection. The RF amplifier from \cite{hoarau2021rf} served as the basis for the design. The signals  were digitised and recorded with oscilloscopes.
\begin{figure}[!htb]
\begin{center}
\includegraphics[width=0.44\columnwidth]{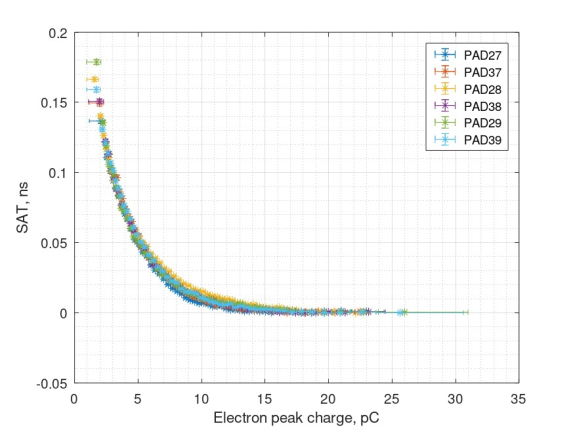}
\includegraphics[width=0.44\columnwidth]{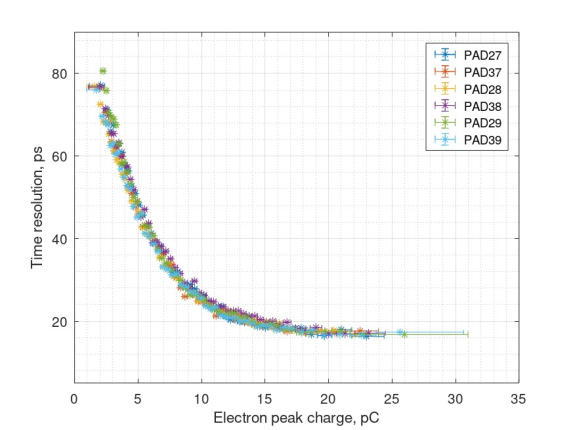}
\caption{Scan measurement of 6 pads: SAT (left) and time resolution (right) for different e-peak charge bins. A similar behaviour is measured for all pads.}  
\label{fig_timing_scan}
\end{center}
\end{figure}
Measurements on this detector and electronics configuration were performed during two RD51 beam test campaigns in Jul. and Oct. 2022. In Jul., two sets of measurements were taken over a four-pad area, while in Oct. a six-pad measurement was performed. The motivation was to study the single-pad time response and signal sharing over multiple pads. The RMS of SAT values for tracks passing through the 4~mm x 4~mm pad central area during the scans are summarised in tables \ref{scan4} and \ref{scan8}. SAT distributions of all measured pads show RMS values of 17.1~ps on average.  To check whether the uniformity of the drift field is maintained, the dependence of the SAT and the time resolution on the e-peak charge was also investigated. A similar behaviour to the prototype with the standard drift gap detector is found, as shown in figure \ref{fig_timing_scan}.

\begin{table}

\parbox{.49\linewidth}{
\centering
\footnotesize
\caption{\footnotesize{RMS of the SAT distributions for 2 x 4 pad scan measurements. Cathode/anode voltage: -465~V/+275~V.}}
\tabcolsep=0.11cm 
\begin{tabular}[t]{l|cccccccc}

PAD & 23 & 24 & 33 & 34 & 27 & 28 & 37 & 38  \\
\hline
RMS [ps] & 17.4 & 17.1 & 17.0 & 17.9 & 17.2 & 16.4 & 17.1  & 16.6 

\end{tabular}
\label{scan4}

}
\hfill
\parbox{.46\linewidth}{
\centering
\footnotesize
\caption{\footnotesize{RMS of the SAT distributions for 6 pad scan measurements. Cathode/anode voltage: -460~V/+275~V.}}
\tabcolsep=0.11cm 
\begin{tabular}[t]{l|cccccc}
 PAD & 27 & 37 & 28 & 38 & 29 & 39 \\
 \hline
 RMS [ps] & 17.7  & 17.3  & 17.1  & 17.1 & 16.4  & 16.9

\end{tabular}
\label{scan8}
}

\end{table}

\section{Conclusions}
\label{sec:6}
The aim of developing a new 100-channel PICOSEC MM detector with an area of 100 cm$^2$ was to achieve a flat and uniform drift gap that would ensure uniform timing over a large area. A prototype based on an embedded ceramic MM board confirmed the uniformity and preserved the time resolution of the single-channel detector. The drift gap thickness was reduced to 180 $\mu$m and timing tests with a newly developed custom RF amplifier were conducted. A timing method for characterising larger areas was implemented and used for measurements. Scans of 10 different pads with a readout chain based on custom amplifiers and oscilloscopes show excellent time resolution averaging 17~ps for all measured pads and the first results show uniformity of the detector with shorter drift gap. New developments are underway to improve stability and robustness and to develop a full readout chain with a multi-channel SAMPIC digitizer \cite{RobustPico}.

\acknowledgments
We acknowledge the financial support of the EP R \& D, CERN Strategic Programme on Technologies 168 for Future Experiments; the RD51 collaboration, in the framework of RD51 common projects; the 169 Cross-Disciplinary Program on Instrumentation and Detection of CEA, the French Alternative 170 Energies and Atomic Energy Commission; the PHENIICS Doctoral School Program of Université 171 Paris-Saclay, France; the Fundamental Research Funds for the Central Universities of China; 172 the Program of National Natural Science Foundation of China (grant number 11935014); the 173 COFUND-FP-CERN-2014 program (grant number 665779); the Fundação para a Ciência e a 174 Tecnologia (FCT), Portugal (CERN/FIS-PAR/0005/2021); the Enhanced 175 Eurotalents program (PCOFUND-GA-2013-600382); the US CMS program under DOE contract 176 No. DE-AC02-07CH11359.



 \bibliographystyle{JHEP}


\end{document}